# Positive Pressure Testing Booths Development and Deployment In Response To The COVID-19 Outbreak

**Kevin Aroom**
Robert E. Fischell Institute for Biomedical Devices
Department of Bioengineering
University of Maryland
College Park, MD 20742
Email: karoom@umd.edu

**Jiawei Ge**
Laboratory for Computational Sensing and Robotics
Department of Mechanical Engineering
Johns Hopkins University
Baltimore, MD 21218
Robert E. Fischell Institute for Biomedical Devices
Department of Mechanical Engineering
University of Maryland
College Park, MD 20742
Email: jge9@jhu.edu

**Lidia Al-Zogbi**∗
Laboratory for Computational Sensing and Robotics
Department of Mechanical Engineering
Johns Hopkins University
Baltimore, MD 21218
Robert E. Fischell Institute for Biomedical Devices
Department of Mechanical Engineering
University of Maryland
College Park, MD 20742
Email: lalzogb1@jhu.edu

**Marcee White**
Total Health Care, Inc.
1501 Division Street, Baltimore, MD 21217
Email: mwhite2@totalhealthcare.org

**Adrienne Trustman**
Health Care for the Homeless
421 Fallsway, Baltimore, MD 21202
Email: atrustman@hchmd.org

**Adena Greenbaum**
STI/HIV Program
Baltimore City Health Department
Baltimore, MD 21202
Email: Adena.Greenbaum@baltimorecity.gov

**Jason Farley**
The REACH Initiative
Johns Hopkins University School of Nursing
Baltimore, MD 21205
Email: jfarley1@jhu.edu

**Axel Krieger**
Laboratory for Computational Sensing and Robotics
Department of Mechanical Engineering
Johns Hopkins University
Baltimore, MD 21218
Robert E. Fischell Institute for Biomedical Devices
Department of Mechanical Engineering
University of Maryland
College Park, MD 20742
Email: axel@jhu.edu

∗Address all correspondence to this author.

## ABSTRACT

*The COVID-19 pandemic left an unprecedented impact on the general public health, resulting in thousands of deaths in the US alone. Nationwide testing plans were initiated to control the spread, with drive-through being the currently dominant testing approach, which, however, exhausts personal protective equipment supplies, and is unfriendly to individuals not owning a vehicle. Walk-up testing booths are a safe alternative, but are too prohibitively priced on the market to allow for nationwide deployment. In this paper, we present an accessible, mobile, affordable, and safe version of a positive-pressure COVID-19 testing booth. The booths are manufactured using primarily off-the-shelf components from US vendors with minimized customization. The booths' mobility allows them to be easily transported within local communities to test a larger subset of the population with fewer transportation options. Moreover, the final bill of materials does not surpass USD 3,900, which is about half of the market price. The booths are air conditioned and HEPA filtered to offer healthcare providers a safe and comfortable working environment. The prototype passed required pressure and air exchange tests, and was positively reviewed by two healthcare professionals. Currently, five booths are deployed and used at the Johns Hopkins University School of Nursing, Baltimore City Health Department, and two community health centers in Baltimore. Our design facilitates walk-up testing in the US, as it decreases PPE consumption; reduces the risk of infection; and is accessible to lower-income communities and non-drivers.*

**Nomenclature**

$P_{cutoff}$  Cutoff pressure in Pa.
$PM_{10}$  Particulate matter. Mass per cubic meter of air of particles with a diameter generally less than 10 micrometers.
$NC_{10}$  Particulate number concentration. Number per cubic meter of air of particles with a diameter generally less than 10 micrometers.
$PM_{2.5}$  Mass per cubic meter of air of particles with a diameter generally less than 2.5 micrometers.
$NC_{2.5}$  Number per cubic meter of air of particles with a diameter generally less than 2.5 micrometers.

## 1 Introduction

As of January 10, 2021, a novel coronavirus (SARS-CoV-2) emerging in December 2019 (commonly known as COVID-19) has resulted in a total of 21,102,069 confirmed cases in the United States of America (US), with 371,084 deaths across age groups [1]. The virus is believed to spread from person to person through respiratory droplets and aerosols [2], with the primary morbidity and mortality linked to pulmonary involvement. The substantial spread and significant impact of the COVID-19 pandemic is ongoing as reports in the US have frequently surpassed 20,000 daily new cases since July 1, 2020 [1]. With effective vaccines only recently developed, and expected distribution to be limited in the coming months to a year, the importance of diagnostic testing and contact tracing remains paramount for protecting the population. Upper respiratory swab specimens, particularly nasopharyngeal and nasal swab specimens, are generally collected for COVID-19 testing. During the collection, a cotton swab is gently inserted through a nostril to the nasopharynx or anterior nares, rotated a few times, and withdrawn. The swab is then placed in a sterile transparent vial, and transported for viral tests (Fig. 1).

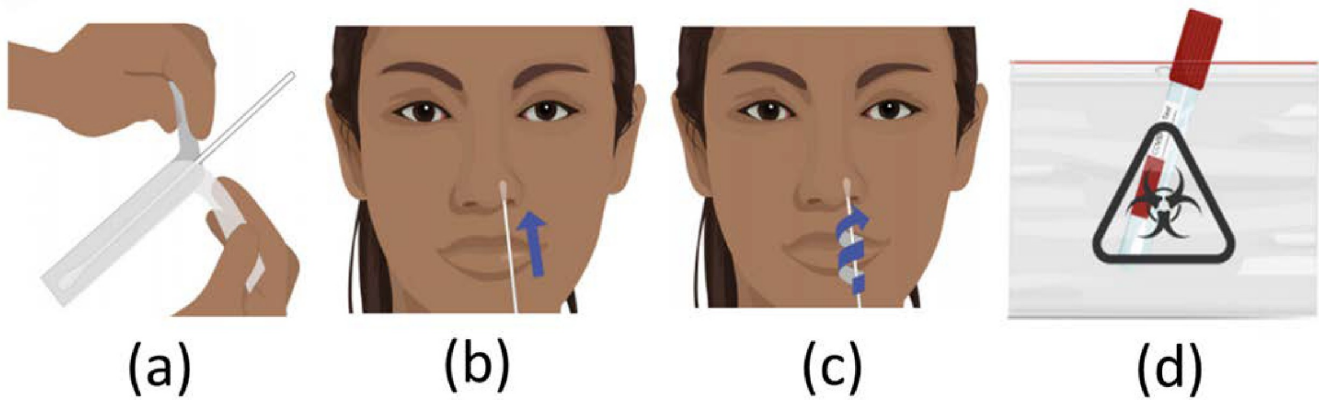

Fig. 1. Representative images of the anterior nasal swab procedures. a) Preparing for swab sample collection; b) Insert the swab into nostril; c) Gently twist the swab while rubbing the anterior nares multiple times; d) Replacing the swab in a vial, and then in a biohazard bag for analysis. Source from cdc.gov/coronavirus.

The US successfully initiated nationwide COVID-19 testing programs, among which are drive-through tests designed to mitigate potential cross-infection at hospitals and clinics, whereby testing sites are commonly set up in parking lots near health centers and select pharmacies [3]. During specimens collection, healthcare providers are required to wear disinfected personal protective equipment (PPE), while the patient remains safely seated inside a vehicle. As of January 10, 2020, the Centers for Disease Control and Prevention (CDC) reported more than 250 million tests reported nationwide [1]. However, drive-through tests impose several limitations. First, the need for replacing all PPE after each test can quickly turn into a burden considering its increasing cost and scarcity. Second, although creating a safe environment for patients, drive-through testing facilities can compromise the healthcare providers' protection. Working in outdoor conditions in uncomfortable positions while wearing multiple layers of PPE can induce fatigue and sample handling errors, and increase the risk of self-contamination during PPE donning and doffing procedures between every patient [4]. Third, drive-through tests are only accessible to a limited portion of the population, namely consisting of people who own vehicles. For instance, multiple studies have shown that African Americans form the highest percentage of COVID-19 cases and deaths [5, 6, 7, 8, 9]. The reasons behind such disparities are linked to the long-standing health and social inequities, as explained by the CDC [10]. Factors such as absence of insurance coverage; not owning a vehicle; placement of testing locations in more affluent areas; and social stigma act as a deterrent for equal testing opportunities. These elements drive the need to propose a solution that allows expansion of testing to areas where under-represented communities can receive convenient access to tests, while offering protection to both medical personnel and patients with limited PPE consumption.

H Plus Yangji Hospital in Seoul, South Korea, pioneered the concept of a walk-up testing facility to accommodate non-drivers [11, 12], and mitigate the absence of space for car queues in dense urban regions. The first testing facilities consisted of a row of depressurized plastic booths the size of a phone booth, with an intercom system and gloves protruding into the booth to allow medical personnel to collect swab samples from the patients. Although daily testing capacity increased from 10 to more than 70 patients, disinfecting internal surfaces while allowing sufficient time for ventilation between each patient proved to be time consuming [11]. An upgraded variant was introduced, whereby the healthcare provider stands inside a positive pressure booth to test the patient on the outside [10, 13]. The cost of purchasing these positive pressure booths amounted to USD 7,500, excluding any shipping rates. Such exorbitant prices can extend far beyond the financial capacity of numerous institutions in the US. A simplified and affordable booth was subsequently designed by the Brigham and Women's Hospital (Boston, MA, USA) [14], yet without making it fully enclosed by leaving the backside of the booth exposed to open air, thus putting the tester at a theoretical risk of infection. A few manufacturing facilities in the US followed in South Korea's footsteps and designed their own positive pressure testing booths (e.g. Garmat, Englewood, CO, USA), but their products are within the same price range as the Korean booths.

In this paper, we propose a positive-pressure testing booth for collecting COVID-19 specimens, with our main contributions attributed to the booth's simple "DIY" (do-it-yourself) design, and the successful clinical deployment and usage around testing facilities in Baltimore, MD, USA. The booth is affordable for local governments, and is designed in a way that facilitates transportation across multiple communities. It is deemed to be safe for healthcare providers by medical experts, and can be manufactured using exclusively off-the-shelf components from vendors such as McMaster Carr, Home Depot, and 80/20. Any research institution or local manufacturer is capable of producing a similar booth within one week, guided by our installation instructions. We also show that fine particulate matter was successfully prevented from entering the booth, eliminating the need to use and replace PPE during sample collection from patients.

This paper is organized as follows: a) Materials and Methods, which encompasses the design specifications and major components of the booth; b) Results, whereby the final product is demonstrated and evaluated against the specifications hereof, with reported results from relevant tests and clinical trials; c) Discussion, providing an analysis of the results and possible improvements; and finally d) Conclusions, summarizing key aspects of our work.

## 2  Materials and Methods

The booth's design needs to meet a set of requirements, the most crucial of which relate to the safety of medical personnel and patients. One safety factor in this case corresponds to maintaining a minimum of 8 Pa (0.82 mm H2O) pressure differential within the booth set by the CDC for positively-pressurized protective equipment [15]. Continuous monitoring and a simple visual indicator showing sufficient pressure level within the booth is required for the occupant to observe. The inflow of air should be filtered from contaminants and aerosol to enhance its quality and safety. The CDC has also put forward ventilation guidelines for protective equipment, specifying a minimum of 12 air exchanges per hour for any enclosed environment as a protective measure against airborne environmental microbes [15].

From a practical standpoint, the booths should be first of all affordable to enable different testing facilities to acquire them, especially charities and facilities in lower-income communities whose resources are often more constrained. Second, raw materials need to be locally available to reduce shipping time and associated costs. Lastly, manufacturing the booths should not require overly expensive equipment, advanced manufacturing facilities and profuse amounts of custom-made parts, which would negatively impact their affordability and public access.

From a mechanical standpoint, the gloves - the only physical medium through which healthcare providers can interact with patients - need to offer sufficient dexterity for handling sample collection and packaging. The gloves should also be comfortably reachable for medical staff with varying height, and the overall dimensions of the booth should accommodate individuals across different heights and body composition. Disinfecting the booth is imperative, therefore its constitutive materials (particularly the gloves) must be resistant to cleaning chemicals, most commonly containing ethanol, hydrogen peroxide and quaternary ammonium [16].

Additional desirable specifications involve the integration of a communication system between healthcare personnel and patients, and a secure and comfortable setup for the patient outside the booth, since swab sample collection is a fairly uncomfortable procedure that can trigger an undesirable physical response. Enabling the patient to visualize the healthcare worker's face can also enhance the patient's comfort. Lastly, the booths should be easily transportable to expand testing to multiple locations, and generally provide sufficient flexibility for repositioning them when needed.

These requirements are summarized in the first column of Table 1, which is also used to oversee the final product's degree of compliance with them. The booth design was thus broken down into 3 major elements; a) the mechanical aspect of the design, b) the ventilation and air conditioning setup, and c) the electrical and circuitry integration, details of which are provided in the subsequent subsections.

**2.1 Mechanical Design**

The proposed testing booth was designed to accommodate a single healthcare provider at a time, as demonstrated in Fig. 2. The overall exterior dimensions of the booth are 1 m × 1.19 m × 2.05 m (width × depth × height), providing ample space for the provider on the inside to feel comfortable and at ease. The frame was constructed using aluminum extrusions due to their lightweight properties, and structural brackets along with concealed anchors were used to reinforce the structure and hold it together. Extrusions were added around the middle of the booth, connecting adjacent vertical members for strengthening purposes. Depending on the environment in which the booths need to be deployed, a steel tread plate or a plastic solid deck pallet were used as a flooring. The steel tread plate is more suited for flat terrains, whereas the pallet can be used for deploying the booth in more rugged areas. The pallet platform raised the booth's overall height to 2.22 m. Handles were added to the external extrusions to assist with the transportation.

Table 1. Summary of the design requirements for the booths and details on the final specifications.

| Design Requirements | Notes and Comments |
|---|---|
| ✓ Minimum +8 Pa internal pressure | The effective pressure inside the booth surpasses +19.6 Pa, corresponding to a safety factor of 2.5. |
| ✓ Continuous monitoring of pressure | A pressure sensor was installed inside the booth to monitor all pressure changes and alert medical personnel in the event of a drastic pressure drop. |
| ✓ Filtered air inflow | Air is channeled inside the booth through a HEPA filter, which was shown to filter out particulate matter of diameter smaller than 10 $\mu m$. |
| Minimum 12 air exchanges per hour | The air conditioning and fan system offer 50 air exchanges per hour. |
| ✓ Affordability | The total cost of the booth (excluding labor cost) amounted to USD 3,900, almost twice as cheap as commercially available booths. |
| ✓ Locally available raw materials | All raw material can be purchased from local vendors. |
| ✓ Manufacturability | No special equipment is needed to manufacture the booth. Any machine shop would suffice. |
| ✓ Materials resistant to decontamination chemicals | All materials are resistant to the most common decontaminants. |
| ✓ Dexterity of gloves | Gloves allow medical personnel to comfortably open a vial, collect a swab sample, and replace it in a vial without additional assistance. |
| ✓ Comfort for various body height and habitus | The booth offers sufficient space for all body types, and a height-adjustable stepper can make the glove ports more reachable to those in need. |
| ✓ Communication system between medical personnel and patient | An intercom system ensures a clear communication between medical personnel and patients. |
| ✓ Tip-proof setup for patients | A chair for patients was designed to prevent any tipping during swabbing. |
| ✓ Transportability | The booths are easily transportable using either four triangular dollies or a pallet jack. |

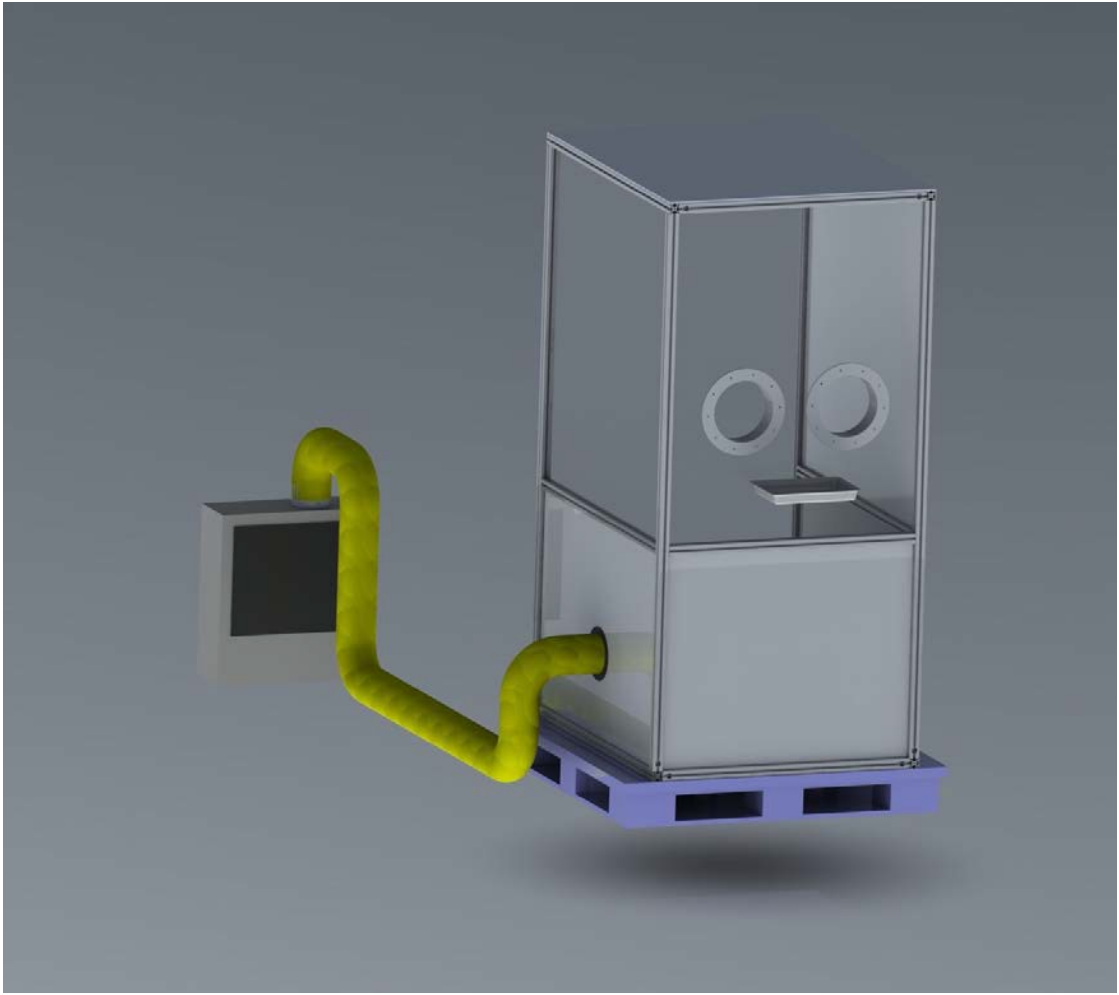

Fig. 2. Computer Aided Design model of the positive-pressure testing booth design.

The roof, lower front, and lower side panels were made of white opaque high density polyethylene (HDPE), whereas the upper side panels, upper front, and door panels were made of clear acrylic. HDPE provided sufficient protection from direct sunlight, whereas acrylic offered healthcare providers the convenience of observing their surroundings. To reduce the potential heating up of the booth from sunlight, and possible visual discomfort from sun rays during the swabbing procedure, the upper side panels and door panels were tinted using an adhesive window tint film. This also offers additional privacy to the medical staff, adding an extra layer of comfort working inside the booth.

To ensure the booth is properly sealed to meet minimum requirements for medical positive pressure of 8 Pa [15], the roof was tightly clamped along its perimeter to the aluminum extrusions using L-shaped steel rails, and the door was taped with weatherseal foam to close off any gaps. A heavy duty rubber mat was also clamped in between the floor and base frame of the booth for a better seal, which also added comfort to the medical personnel.

A pair of Hypalon gloves (8Y1532/9Q, Honeywell, Charlotte, NC, USA) were attached to two flange ports, with a specimen collection tray mounted on the outside beneath them. The gloves are resistant to chemical contamination, as well as standard chemical disinfecting procedures. A height adjustable platform (0.1, 0.15, and 0.2 m) was placed inside the booth to allow healthcare providers of different heights to comfortably reach the glove ports. Since healthcare providers usually work with tablet computers or laptops, a side table was mounted on the inside of the booth for their convenience. Patients are tested while sitting on a chair to offer better control and maneuverability to the medical staff. Since nasopharyngeal swab collection can be somewhat uncomfortable, and patients tend to lean backwards during the procedure, a tip-proof chair setup was designed to avoid possible injuries. The chairs were clamped onto a wooden platform covered by an HDPE sheet for easier cleaning, some of which were elevated using raisers to compensate for the added height from the pallets.

**2.1.1 Ventilation and Air Conditioning Design**

Generating a positive pressure environment inside the booth requires a forced air system capable of delivering air at a specific flow rate with expected backpressure. To this end, a portable air conditioning (AC) and filtering system were installed. The AC (T9F796000, Global Industrial, Port Washington, NY, USA) was integrated with a booster fan (OA172AP- 11-1TB, DigiKey, Thief River Falls, MN, USA), concurrently supplying a maximum of 899 m3/hr of airflow. The fans can be turned on independently, or in combination with the cooling option. The cooling option only serves the purpose of moderating the temperature and reducing humidity levels inside the booth during summers, as the fan by itself is capable of providing sufficient airflow to create a positive pressure. To ensure a safe environment, the airflow was channeled through a HEPA filter (HEPA 500 F321, Dri-Eaz, Burlington, WA, USA) that was mounted inside the booth. The filter, listed to block 99.97% of 0.3 $\mu$m oily aerosol particles, was mounted on the booth, and sealed using silicone on its periphery to prevent unfiltered air leakage. A pressure sensor (SDP810-125PA, Sensirion AG, Switzerland, USA) controlled by an Arduino Nano Every board (Arduino, Somerville, MA, USA) was mounted inside the booth for monitoring pressure changes. The board was placed inside an electrical box, and connected to a light emitting diode (LED) (PML50RGFVW, Visual Communications Company, Carlsbad, CA, USA) indicator to warn healthcare providers about an unexpected pressure drop, which can result from an improperly sealed door, dirty filter, or an air leak from mechanical damage to the booth. The accuracy of the pressure sensor is listed as $\pm 3\%$. The CDC requirement for a safe positive pressure environment is 8 Pa, therefore the readings of said pressure can range between 7.76 and 8.24 Pa. A conservative safety factor of 2.5 was considered for the measurements, therefore a cutoff pressure of +19.6 Pa was used to account for sensor inaccuracies, below which the booth is deemed to be unsafe for regular operation. The booth is intended to be used exclusively under positive pressure conditions. The pseudo-code for the sensing mechanism is presented below.

---

**Algorithm 1** Pressure Sensing

---
1: **input:** cutoff pressure $P_{cutoff}$ = 19.6 Pa
2: **procedure**
3:     initialize previous pressure reading → 0
4:     **void main()**
5:         collect current pressure reading
6:         **if** current pressure reading < $P_{cutoff}$ **then**
7:             LED turns RED
8:         **else**
9:             **if** previous pressure reading < $P_{cutoff}$ **then**
10:                 LED turns OFF
11:                 wait for 10 seconds
12:             **else**
13:                 LED turns GREEN
14:     update previous pressure reading → current pressure reading
15: **output:** LED color change

---

**2.2 Electrical Design**

The booth requires 120V AC power to operate as designed. A single hospital-grade plug is used to make the connection to main power. The booth is equipped with receptacles to provide additional convenience to the healthcare providers. A weatherproof external outlet is used to supply a power connection for external items such as the ventilation system. A dimmable LED light (FP2X2/4WY/WH/HD, Commercial Electric Products, Cleveland, OH, USA) was installed on the ceiling to allow testing even at later times of the day. The pressure sensor was also powered from the booth, and an additional charging outlet was installed on the inside. To ensure a clear communication between the medical personnel and patients, a hands-free two-way intercom system was installed. The intercom (TW102, Retekess Technology Co., Shenzhen, China) consists of a base speaker/microphone placed inside the booth, from which the medical staff has full control over the powering of the device and volume of the outputs, and a substation speaker/microphone placed outside the booth, closer to the patient. The substation was positioned inside a 3D-printed cradle, attached to one of the vertical aluminum extrusions.

## 3 Results

A total of five booths were successfully manufactured and deployed at various testing facilities across Baltimore, MD (Fig. 3): a) Two community health centers; b) Baltimore VA Medical Center; and d) Johns Hopkins University School of Nursing. In the first column of Table 1, a check mark indicates that the requirements set in the Methods section have been met. Indeed, all requirements were satisfied, and additional details on how each specification was met are presented in the second column. The specifications that were fulfilled as well as additional testing results are more thoroughly analyzed in the following subsections.

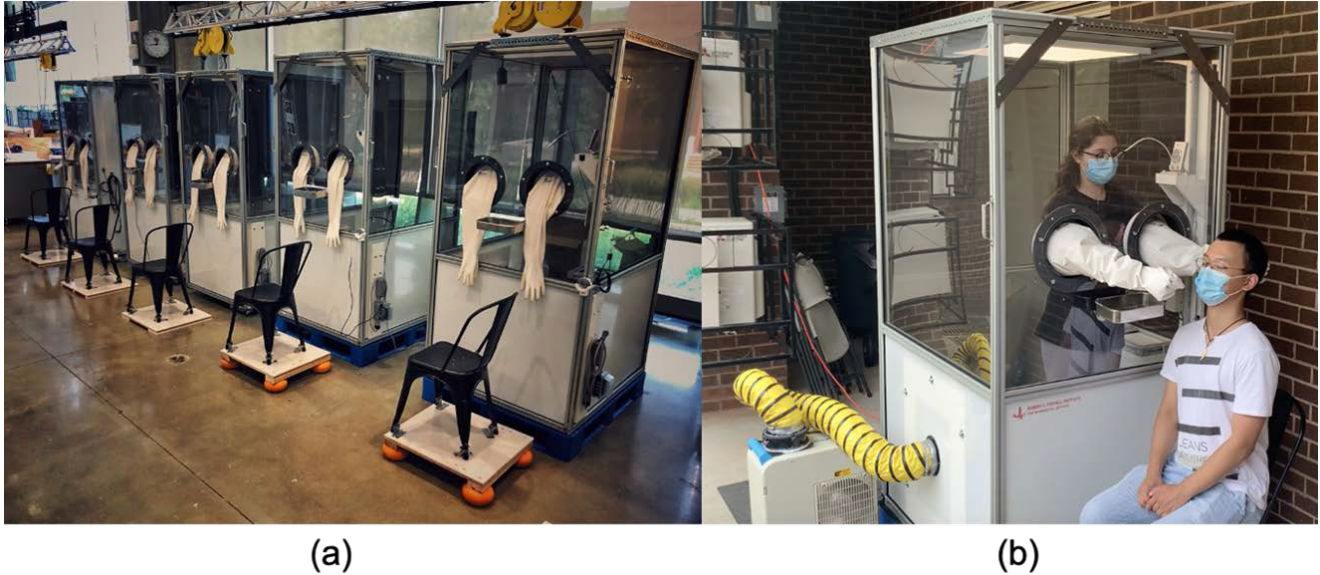

Fig. 3. a) Overview of the five manufactured testing booths; b) A picture of our team members demonstrating the booth usage to healthcare providers at Total Health Care, Baltimore, MD.

### 3.1 Booth Design and Mobility

Two main designs were adopted for the manufacturing process, with the major difference being in the platform on which they were constructed, namely a steel tread platform and a deck pallet. The steel platform facilitates the placement of four triangular dollies at the corners of the booths for easy transportation, whereas the pallet platform allows to move the booth using a pallet jack. All booths can be transported by one person of average body composition, however two people are recommended to ensure the stability of the booth and raise the steel tread plate once for positioning the dollies beneath them.

### 3.2 Safety

To evaluate the safety of the booth, pressure tests and particle filtration tests were conducted. Positive pressure is a critical factor in ensuring medical personnel's safety, since it prevents contaminants and droplets from infiltrating into the booth. In our designs, an external air supply system is responsible for maintaining a positive pressure inside the booth, channeled through a HEPA filter. The air supply can be at ambient temperature (fan only mode), or at a colder temperature (fan and cooling mode). The pressure of the booths was evaluated at both modes, amounting to an average of $28.4 \pm 5.9$ Pa in $9.4 \pm 3.3$ seconds during fan only mode, and in $6.8 \pm 0.8$ seconds when cooling is turned on. The obtained pressure readings for all booths surpassed the minimum threshold of 8 Pa set by the CDC safety requirements for positive pressure rooms [15]. The ventilation rate was evaluated to exceed 50 air exchanges per hour, estimated by interpolating the flowrate of the current system from pressure-flow measurements at zero flow and unrestricted flow. The obtained ventilation rate satisfies the CDC recommendations of 12 air exchanges per hour [15].

We used a particle detection sensor (SPS30, Sensirion AG, Switzerland) to analyze the filtration capacity of the booth. Inhalable respiratory droplets that carry the COVID-19 virus are greater than 5 $\mu$m and less than 10 $\mu$m in diameter, whereas aerosols and droplet nuclei have a diameter smaller than 5 $\mu$m [17, 18]. Therefore, PM10 and NC10 evaluation metrics were chosen to assess each booth's safety in both fan only and fan with cooling modes. PM2.5 and NC2.5 were reported for a more accurate measurement of the fine inhalable particles, since transmission of COVID-19 through aerosol has been shown in [2]. The particle sensor was placed inside each booth, making sure that they were well ventilated prior to being closed, such that the particle number inside matches that of the external environment. The booths were subsequently closed, and the measurement began when the air conditioning and/or fan system was turned on. The results are reported in Fig. 4, whereby it can be observed that after a certain period of time, all booths were completely free from particulate matter. Particulate matter for all booths and all modes reached a constant zero reading after $5.8 \pm 2.3$ min using fan only, and $2.9 \pm 0.7$ min with the cooling. The number concentration of particles exhibited the same behavior, decaying to a steady zero within $6.0 \pm 2.4$ min with a fan, and $3.2 \pm 0.7$ min with fan and cooling. For both pressure and particle tests, the experiments reached a favorable outcome faster with the cooling function on, since it provides additional airflow to that of the fan, resulting in a faster air exchange rate.

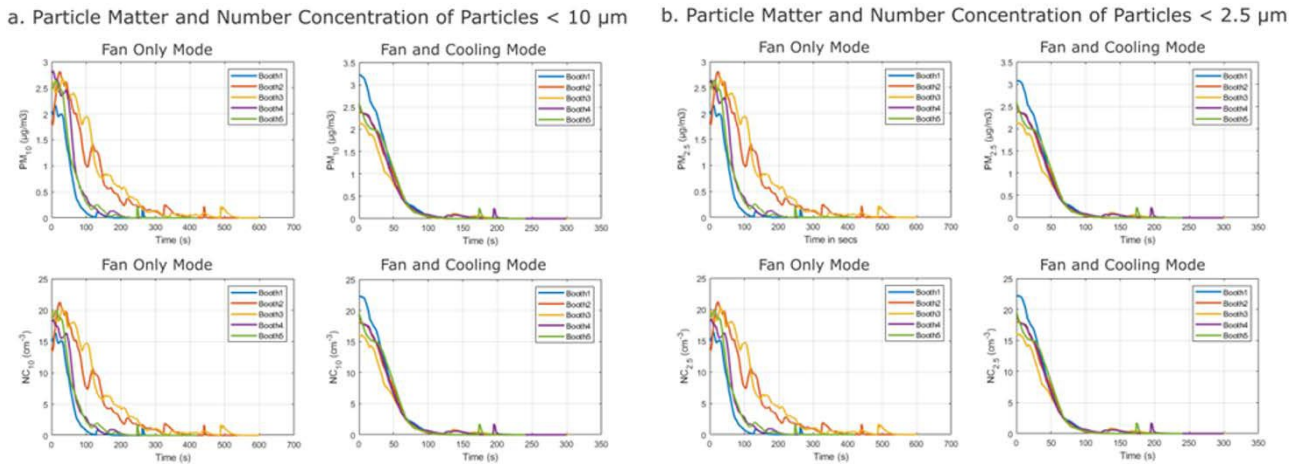

Fig. 4. $PM_{10}$, $PM_{2.5}$, $NC_{10}$ and $NC_{2.5}$ inside the booth when only the fan is operating, and when both fan and cooling are operating.

### 3.3 Cost and Manufacturability

The final bill of materials did not exceed USD 3,900 for any of the booths, which is substantially cheaper than the commercially available ones. Mass production could, of course, significantly reduce the per booth material costs. Additionally, the booths are entirely made of off-the-shelf materials and components, which facilitates the manufacturing process, and gives opportunity to various local institutions and organizations for developing their own. The availability of certain products, however, can depend on the market demand; a temporary shortage of acrylic sheets for instance occurred due to the deployment of sneeze guards across the nation, yet replacement materials (such as polycarbonate sheets) can usually be purchased from local plastic vendors. Some components had to be modified, such as mechanically and electrically integrating the booster fan with the air conditioner. The intercom wiring connecting the substation speaker to the base one also had to be divided into two parts to add an adapter for tightly sealing the corresponding opening on the booths.

### 3.4 Prototype Test

Nasopharyngeal swabs were collected using our first booth prototype to receive feedback from medical personnel for finalizing our design. Eight patients were tested by a nurse practitioner and a physician using the prototype booth. The swab procedure can be divided into five steps, which include an inquiry step during which the healthcare provider communicates with the patient to inform them about the procedure; sample collection using a swab; patient leaving; specimen packaging; and decontamination. The median times required for each step are listed in Fig. 5 in a boxplot. The average times required for each of the aforementioned steps are, respectively: $75.0 \pm 24.6$ seconds; $27.4 \pm 5.8$ seconds; $8.5 \pm 10.2$ seconds; $46.0 \pm 18.6$ seconds; $96.5 \pm 6.8$ seconds. It hence takes an average of 4 min and 23 seconds for one patient to get swabbed. Moreover, after collecting specimens from all eight patients, we collected multiple swab samples from the booth's interior surfaces, the results of which were shown to be negative for COVID-19.

### 3.5 Usability Test

We compiled a set of 9 questions sent out to the healthcare providers who used our booth to collect COVID-19 specimens. The answers were based on a Likert-scaled evaluation, ranging from 1 (extremely dissatisfied) to 10 (extremely satisfied). The goal of the questionnaire was to receive input from the medical personnel for improving the prototype and future versions of the booth by addressing their direct needs. The results are shown in Table 2, demonstrating the overall satisfaction and approval of the healthcare providers. The response of the first healthcare provider is reported individually to highlight the improvement of the final design. The average of 8 providers' responses are listed; these providers have not evaluated the prototype design.

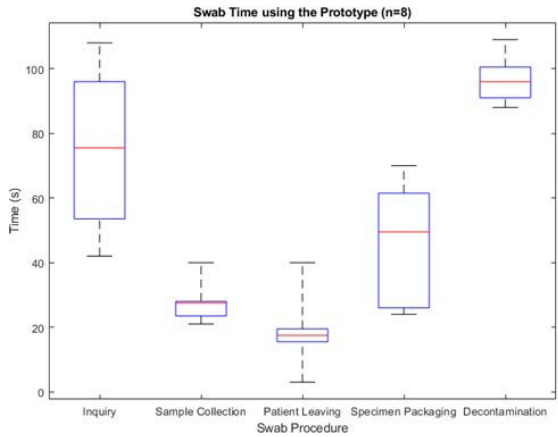

Fig. 5. Box plots of nasopharyngeal swab times for eight volunteer patients throughout the different stages of the swab test.

Table 2. Usability questionnaire results reported on a Likert-scaled evaluation ranging from 1 (extremely dissatisfied) to 10 (extremely satisfied).

|  | Prototype | Final Designs | |
| --- | --- | --- | --- |
|  | Health Provider 1 | Health Provider 1 | Average of 8 health providers responses |
| Comfort | 10 | 10 | 7.87±2.23 |
| Dexterity | 10 | 8 | 6.37±2.13 |
| Safety | 10 | 10 | 10.00±0.00 |
| Communication | 8 | 10 | 6.71±1.80 |
| Temperature comfort | 8 | 9 | 8.12±2.17 |
| Air quality | 6 | 10 | 9.50±1.41 |
| Patient's response | 10 | 9 | 8.50±1.69 |
| Booth dimension | 10 | 10 | 8.50±1.41 |
| Overall Impression | 10 | 10 | 8.50±2.00 |

### 3.6 Clinical Approval

PPE in the US is generally evaluated and regulated by the FDA prior to clinical usage. Due to the COVID-19 public health emergency, some PPE (such as medical gloves), whereby its intended usage is not expected to assist in preventing the transmission of infectious disease, may not require the manufacturer to submit a notice to the FDA prior to marketing their products. Face masks are subject to FDA regulation, however regulatory flexibility was introduced for such products, such as eliminating the 510(k) premarket notification and Quality System Regulations. Respirators, on the other hand, still require the National Institute for Occupational Safety and Health's (NIOSH) approval. The testing booths do fall under the umbrella of devices that prevent a disease's transmission, as well as under the Emergency Use Authorization provided by the FDA, in particular for Face Shields and Barriers. The booths are used at the medical personnel's discretion.

## 4 Discussion

The proposed booths have shown to be a feasible solution for collecting samples from patients in a safer and more cost-effective way. As a walk-up testing means, they are accessible to all communities, especially minorities. They are also affordable, and easy to manufacture and assemble using readily-available products from local vendors. This speeds up the manufacturing process, allowing facilities to quickly meet any rising demand for such testing booths. They are also mobile, which means that they can service more communities depending on the need. Lastly and most importantly, the booths offer a safe working environment to the medical personnel, since our results show that inhalable particles of diameter smaller than 10 $\mu$m are eliminated through the HEPA filter.

The booths were well received by the medical staff at the five deployment sites, three of which are located in facilities serving lower-income communities. The low cost and ease of manufacturability are encouraging factors, and should help local governments deploy the booths in clinics facing challenges in procuring appropriate PPE. This booth has the opportunity to address testing deserts, or areas with low testing access, which are concentrated in lower income communities. As such, this initiative provides an opportunity to increase testing amongst those with disparate risk to greater morbidity and mortality.

Additionally, walk-up booths offer a range of advantages over the prevalent drive-through testing, amongst which is enhanced testing throughput, with the testing time per patient using the walk-up booth amounting to an average of 4 minutes, as compared to 15 minutes using drive-through facilities [10]. The 4-minute testing time during the pilot study is expected to further decrease as clinical teams get into routine workflow. Walk-up booths may offer a more comfortable and air-conditioned working environment to the healthcare providers, reducing handling errors, uncomfortable postures, and mental stress. Positively-pressured testing booths are also a better alternative to negatively-pressured ones, since they require less time and effort to decontaminate and ventilate the workspace, and are safer for patients as they would be tested in a more open environment.

Another advantage of our booths is the flexibility of the design, which leaves room for modification and personalization; the dimensions of the booth, materials used, and additional accessories can be altered depending on the need or preference of the users. With any possible design change, however, it is crucial to maintain an acceptable internal positive pressure, hence special care needs to be taken with the choice of fan and air conditioning units in order to supply suitable airflow. The booths are also designed in a way that allows for easy replacement of components. The HEPA filter, which is susceptible to collecting dust and particles, especially when deployed outdoors in a city, can be easily taken out for cleaning or replacement. The gloves can be replaced, as they are held by a quick-release clamp to the ports. The HDPE and acrylic sheets can also be substituted, however, it would involve a more demanding disassembly procedure. Such interchangeability extends the longevity of the booths, allowing to salvage components, and thus reduce costs in the long run.

Major limitations of the design include the lack of robustness against adverse weather conditions (such as rain and strong wind) and poor overnight outdoor security, as we recommend to place the booths inside protective sheds for storage. Purchasing a large enough shed to house a booth will incur additional costs. If a shed is not available, developing a waterproof design can protect the booth from the weather at an additional cost, but does not solve the security issue. Overnight security at health facilities is required to prevent theft and damage, otherwise the booths have to be moved indoors every night. Additionally, labor cost has not been factored into the final cost of the product as it can vary from one facility to another, meaning that the final cost of a booth might exceed USD 3,900, which can, however, be mitigated by a reduction in material cost when ordered in bulk for larger production runs. Some level of customization is still required (e.g., setting up the electrical connections for the AC and fan), which might appear to be challenging for inexperienced users. Future research directions would involve a clinical analysis of the booth's safety and efficacy in a larger experiment, and over an extended period of time. Samples can be collected from the booth's interior and exterior surfaces on a continuous basis to analyze whether infected patients impose any risk on the medical staff, as well as other non-infected patients. The number of infected medical personnel through the booths (if any), should also be compared to drive-through and other testing methods.

## 5 Conclusion

This paper presents the details of the design and implementation of five positive pressure testing booths for COVID-19 sample collection. The booths are equipped with an external fan and air conditioning system, channeling airflow through a HEPA filter for added safety. The pressure for all booths meets the standard minimum for medical applications, and particle detection tests demonstrated the efficiency of the filter at eliminating particles of sizes smaller than 2.5 and 10 $\mu$m. The booths were positively reviewed by two healthcare professionals, and are currently in use at five different testing facilities across Baltimore, MD.

## 6 Funding

This project was generously supported by the University of Maryland, College Park, the Johns Hopkins Malone Center for Engineering in Healthcare, and the REACH Initiative of the Johns Hopkins University School of Nursing.

## References

[1] Center for Disease Control and Prevention, 2020. "COVID-19 Cases, Deaths, and Trends in the US| CDC COVID Data Tracker". https://covid.cdc.gov/covid-data-tracker.

[2] Fears, A., Klimstra, W., Duprex, P., Hartman, A., Weaver, S., Plante, K., Mirchandani, D., Plante, J., Aguilar, P., Fernández, D., Nalca, A., Totura, A., Dyer, D., Kearney, B., Lackemeyer, M., Bohannon, J., Johnson, R., Garry, R., Reed, D., and Roy, C., 2020. "Comparative dynamic aerosol efficiencies of three emergent coronaviruses and the unusual persistence of SARS-CoV-2 in aerosol suspensions". *medRxiv*, Apr.

[3] U.S. Department of Health and Human Services, 2020. "COVID-19 Testing". https://www.hhs.gov/coronavirus/testing/index.html.

[4] Choi, S., Han, C., Lee, J., Kim, S.-I., and Kim, I. B., 2020. "Innovative screening tests for COVID-19 in South Korea". *Clinical and Experimental Emergency Medicine,* **7**(2), June, pp. 73–77.

[5] Adegunsoye, A., Ventura, I. B., and Liarski, V. M., 2020. "Association of Black Race with Outcomes in COVID-19 Disease: A Retrospective Cohort Study". *Annals of the American Thoracic Society*, July. Publisher: American Thoracic Society - AJRCCM.

[6] Millett, G. A., Jones, A. T., Benkeser, D., Baral, S., Mercer, L., Beyrer, C., Honermann, B., Lankiewicz, E., Mena, L., Crowley, J. S., Sherwood, J., and Sullivan, P. S., 2020. "Assessing differential impacts of COVID-19 on black communities". *Annals of Epidemiology,* **47**, July, pp. 37–44.

[7] Price-Haywood, E. G., Burton, J., Fort, D., and Seoane, L., 2020. "Hospitalization and Mortality among Black Patients and White Patients with Covid-19". *New England Journal of Medicine*, May. Publisher: Massachusetts Medical Society.

[8] Elbaum, A., 2020. "Black Lives in a Pandemic: Implications of Systemic Injustice for End-of-Life Care". *Hastings Center Report,* **50**(3), pp. 58–60. eprint: https://onlinelibrary.wiley.com/doi/pdf/10.1002/hast.1135.

[9] Dyer, O., 2020. "Covid-19: Black people and other minorities are hardest hit in US". *BMJ,* **369**, Apr. Publisher: British Medical Journal Publishing Group Section: News.

[10] Centers for Disease Control and Prevention, 2020. "COVID-19 Health Equity Considerations and Racial and Ethnic Minority Groups". https://www.cdc.gov/coronavirus/2019-ncov/community/health-equity/race-ethnicity.html.

[11] Kim, S. I., and Lee, J. Y., 2020. "Walk-Through Screening Center for COVID-19: an Accessible and Efficient Screening System in a Pandemic Situation". *Journal of Korean Medical Science,* **35**(15), Apr.

[12] Kim, S. I., and Lee, J. Y., 2020. "Letter to the Editor: How Did We Solve the Risk of Cross-Infection after Testing by the Walk-Through System Pointed Out by Many Authors?". *Journal of Korean Medical Science,* **35**(16), Apr.

[13] Bong-geun, S., 2020. Evolving Corona Collection…This time,'Super Speed Walking Through' came out, Mar.

[14] Brigham Health Hub, 2020. "Innovative COVID-19 testing booth". https://brighamhealthhub.org/covid-19/transforming-the-way-clinicians-test-covid-19-patients.

[15] American Psychological Association, 2003. Guidelines for Environmental Infection Control in Health-Care Facilities: (545922006-001). Tech. rep., American Psychological Association.

[16] Mayo Clinic. "Fight coronavirus transmission at home".

[17] Jayaweera, M., Perera, H., Gunawardana, B., and Manatunge, J., 2020. "Transmission of COVID-19 virus by droplets and aerosols: A critical review on the unresolved dichotomy". *Environmental Research,* **188**, Sept., p. 109819.

[18] US EPA, OAR, 2016. "Particulate Matter (PM) Basics". *US EPA*, Apr. https://www.epa.gov/pm-pollution/particulate-matter-pm-basics.